\documentclass[useAMS,usenatbib]{mn2e}
\usepackage{latexsym,amsmath}
 \usepackage[dvips]{graphicx}
 \usepackage{natbib}

\title[Miranda's dynamical history]
  {A numerical exploration of Miranda's dynamical history}
\author[E. Verheylewegen, B. Noyelles and A. Lemaitre]
  {E.~Verheylewegen\thanks{emilie.verheylewegen@fundp.ac.be~(EV)}, 
  B.~Noyelles\thanks{benoit.noyelles@fundp.ac.be~(BN)}
  and A.~Lemaitre\thanks{anne.lemaitre@fundp.ac.be~(AL)},\\
  NaXys,
      University of Namur,
      8, Rempart de la Vierge, B-5000 Namur, Belgium\\}
     
\date{Released 2012 Xxxxx XX}

\pagerange{\pageref{firstpage}--\pageref{lastpage}} \pubyear{2002}

\def\LaTeX{L\kern-.36em\raise.3ex\hbox{a}\kern-.15em
    T\kern-.1667em\lower.7ex\hbox{E}\kern-.125emX}

\begin{document}

\label{firstpage}

\maketitle

\begin{abstract}
 The Uranian satellite Miranda presents a high inclination ($4.338^\circ$) and evidences of resurfacing. It is accepted since 20 years (e.g. Tittemore and Wisdom 1989, 
Malhotra and Dermott 1990) that this inclination is due to the past trapping into the 3:1 resonance with Umbriel. These last years there 
is a renewal of interest for the Uranian system since the Hubble Space Telescope permitted the detection of an inner system of rings and 
small embedded satellites, their dynamics being of course ruled by the main satellites. For this reason, we here propose to revisit the long-term 
dynamics of Miranda, using modern tools like intensive computing facilities and new chaos indicators (MEGNO and frequency map analysis). 
As in the previous studies, we find the resonance responsible for the inclination of Miranda and the secondary resonances associated, 
likely to have stopped the rise of Miranda's inclination at $4.5^\circ$. Moreover, we get other trajectories in which this inclination reaches $7^\circ$. 
We also propose an analytical study of the secondary resonances associated, based on the study by Moons and Henrard (1993).
\end{abstract}


\section{Introduction}

\par In the 1980s, the Voyager 2 spacecraft gave us a better knowledge of Uranus and its satellites (see e.g. \citealt{smith86}). 
It revealed in particular that Miranda and Ariel have been resurfaced. Moreover, orbital models compared to astrometric observations showed that Miranda had a 
significant inclination $i_M$,  of the order of $4.338^{\circ}$ among GUST86\footnote{General Uranus Satellite Theory}\citep{laskar86}. These facts induced several dynamical studies in the late 1980s and early 
1990s \citep{dermott88,malhotra88,tittemore89,malhotra90,tittemore90,henrard90} showing that the current inclination of Miranda is probably due to a former 3:1 resonance 
with Umbriel. Indeed, tidal interactions with Uranus are supposed to push the satellites outward, meeting orbital resonances. Once the system is trapped into this 3:1 resonance, the inclination of Miranda 
is pumped and the amplitude of libration of the resonant argument rises as the trajectory meets several secondary resonances. The capture in one of these resonances leads the trajectory to the edge of the primary resonance
involving the exit of this latter. Eventually, the inclination of Miranda ceases to be pumped and the 2~satellites restart to migrate independently of each other. This scenario of evolution has been intensively studied by the above authors. With this work, we confirm the ideas developed in the past and introduce modern numerical 
tools to update the problem.

\par The reason is that there is a renewal of interest these last years for the Uranian system. First, the Hubble Space Telescope allowed the 
discovery of a whole system of rings and inner satellites (see e.g. \citealt{showalter06}), whose dynamics is of course widely influenced by the main satellites. 
These inner satellites present interesting dynamical configurations and mysteries, as for instance the poorly understood dynamics of Mab \citep{kumar11} or the instability 
of Cupid, Belinda, Cressida and Desdemona on a time scale of $10^3-10^7$ years \citep{french12}. Secondly, a new scenario has recently been proposed by \citet{boue10} to explain the huge obliquity of Uranus ($\approx98^{\circ}$). 
This scenario involves a former giant satellite, whose gravitational torque was strong enough to tilt the planet \citep{morbidelli2012}. We can also mention the work of \citet{deienno11} showing that the presence of the current main satellites is consistent with 
the migration of Uranus as predicted by the Nice model. Finally, Uranus and its satellites are the target of the proposal of a space mission Uranus Pathfinder \citep{arridge12}. 
For all of these reasons, we propose to revisit the dynamical history of the main Uranian satellites with powerful numerical means and tools, 
such as frequency analysis \citep{laskar93} and the MEGNO chaos indicator \citep{cincotta00}.

\par This article is split into the following sections. First the key features of the system are summarised (cfr. Section~\ref{S:key_features}). We present the system and the 
mean motion resonance 3:1 between Miranda and Umbriel that the system should have encountered in its history. 
The details of the initial conditions used throughout this study are given in order to group
the information needed to redo the numerical analysis. Secondly, based primarily on the work of \citet{tittemore88} and \citet{malhotra90}, we present the 
full Hamiltonian (cfr. Section~\ref{S:P1}) and its averaged form (cfr. Section~\ref{S:P2}). These two sections remind the description of the system via Hamiltonian formalism. 
The introduction of canonical variables as Delaunay variables allows us to develop a perturbative theory and
to obtain an Hamiltonian with two degrees of freedom (angle-action) by averaging over the fast angles. We also introduce the equations related to the tidal effects in the two models.
\par We then present the new numerical tools implemented to study the system. The numerical integrations of the full 
system on a sufficient large time scale were a major problem in the previous studies. We perform numerical integrations over $1$ Myr on the system of Uranus with its five main satellites and confirm previous results. 
We use numerical tools like the chaos detector MEGNO to represent the global dynamics of the system (cfr. Section \ref{S:Numerical_Methods}). 
We also improve the resolution and the details of the maps by studying the variations of the orbital elements involved in the mean motion resonance. These scales are introduced in the section \ref{S:Numerical_Methods}
and the results on the system are summarised in the section \ref{S:Results}. These new tools allow us to extend previous studies by the introduction of new visualisations 
of the phase plane of the considered problem. We show that the combination of numerical methods like chaos maps and frequency analysis (cfr. Section \ref{S:Results}) allows 
to detect some particular behaviours of the system. We find two regions surrounding the center of libration where two secondary resonances are superimposed. The choice of the initial
condition for a trajectory is then primordial for the future of the system: the combination of the two secondary resonances tends to increase the chaos and to result in
another scenario than the exit at $4.5^\circ$ for the inclination of Miranda. This is a part of our conclusions and perspectives presented in the last 
section \ref{S:conclusionsandperspectives}.

\section{Key features of the system}
\label{S:key_features}
The system of Uranus has five main satellites, from closest to farthest with respect to Uranus; Miranda, Ariel, Umbriel, Titania and Oberon. Although these satellites are not currently locked into 
orbital resonances, there are many clues indicating probable passages through mean motion resonances in the past: we observe important resurfacings
of Miranda and Ariel and some abnormalities in the current orbital elements. In view of these orbital elements of the satellites, in
particular the high inclination of Miranda (cfr. Table \ref{orbital_elements_satellites}), we analyse a mean motion resonance acting on the inclinations. The first one
encountered by the system in the past is the mean motion resonance Miranda-Umbriel 3:1 defined by 6 possible resonant arguments:\\

\begin{tabular}{@{}ll@{}}
 $\theta_1=\lambda_M-3\lambda_U+2\Omega_M$&\quad\quad $[I_M^2]$   \\
 $\theta_2=\lambda_M-3\lambda_U+\Omega_M+\Omega_U$&\quad\quad $[I_MI_U]$  \\
 $\theta_3=\lambda_M-3\lambda_U+2\Omega_U$&\quad\quad $[I_U^2]$ \\
 $\theta_4=\lambda_M-3\lambda_U+2\varpi_U$&\quad\quad $[e_U^2]$ \\
 $\theta_5=\lambda_M-3\lambda_U+\varpi_M+\varpi_U$&\quad\quad $[e_Me_U]$ \\
 $\theta_6=\lambda_M-3\lambda_U+2\varpi_M$&\quad\quad $[e_M^2]$  \\
\end{tabular}

\vspace{0.4cm}
\noindent
where, in the left column, $\theta_i$ are the resonant arguments for the primary resonances with $\lambda_i$, the mean longitudes, $\Omega_i$ the ascending nodes and, $\varpi_i$ the pericenters. 
The indices $M$ and $U$ stand respectively for Miranda and Umbriel. The right column is the type of the resonance and corresponds
 to the first non-zero term associated with the cosine of the 
angle $\theta_i$ in the perturbative potential (cfr. Eq.\ref{S1.Eq_potgrav} in the following section).

These six primary resonances are not well separated because the oblateness of Uranus, further indicated by
the parameter $J_2$,
is rather small (cfr. Table \ref{physical_parameters_Uranus}) which results that the isolated resonance theory is only applicable for small eccentricities and inclinations 
(\citealt{dermott88}; \citealt{tittemore88}; \citealt{malhotra90}). 
When the inclination of Miranda increases, this classical theory breaks down: it has been shown that the proximity of other primary resonances implies the exit of that resonance 
\citep{tittemore90} due to a commensurability between the libration frequency of the resonant argument and the circulation frequency of a close primary resonance,
in other words, due to the passage in a secondary resonance zone (\citealt{tittemore90}; \citealt{malhotra90}).

Our main purpose being the study of the high inclination of 
Miranda, we focus on the primary resonance of type $I_M^2$. The mixed primary resonance $I_MI_U$ implies also the rise of the inclination of Umbriel,
which is currently in the same order of the inclinations of the three other satellites Ariel, Titania and Oberon. For this reason, we do not focus on this type of resonance.
\par For the numerical analysis of the system, we use as initial conditions the values given by the JPL\footnote{http://ssd.jpl.nasa.gov}. All the parameters used are gathered 
in Table \ref{orbital_elements_satellites} for the mean orbital elements of the satellites, in Table \ref{physical_parameters_satellites} for the physical parameters of the 
satellites and, in Table \ref{physical_parameters_Uranus} for the physical parameters of Uranus.
\par The initial conditions in our numerical experiments are fixed to the current values, except for the inclination of the satellite Miranda and the ratio
of semi-major axes of the two satellites Miranda and Umbriel. Considering tidal evolution, we modify
these elements to reproduce the capture in the mean motion resonance 3:1 and to study the consequences on the system. Since Miranda is the closest main satellite of Uranus, we
also use its periods to determine the integration steps. All these points are detailed in each numerical experiment presented in the following sections.

\begin{table*}
 \centering
 \begin{minipage}{140mm}
\caption{Mean orbital elements of the five main satellites \citep{laskar87}: $a$ is the semi-major axis, $e$ the eccentricity, $\omega$ the pericenter, $M$ the mean anomaly, $i$ the inclination, $\Omega$ the ascending node, 
$n$ the mean motion. The variables $P$ and $P_{\Omega}$ stand for the orbital and the node periods respectively.}
\label{orbital_elements_satellites}
\begin{tabular}{@{}lccccccccc@{}}
\hline
 Satellites & $a$ & $e$ & $\omega$ & $M$ & $i$ & $\Omega$ & $n$ & $P$& $P_{\Omega}$\\
& (km)&& (deg)& (deg)& (deg)& (deg)&(deg/day)&(days)&(yr)\\
\hline
Miranda  & $129\ 900$ & $0.0013$ &  $68.312$ & $311.330$ & $4.338$ & $326.438$ & $254.6906576$ & $2.520$ & $17.727$ \\
Ariel    & $190\ 900$ & $0.0012$ & $115.349$ &  $39.481$ & $0.041$ &  $22.394$ & $142.8356579$ & $4.144$ & $57.248$ \\
Umbriel  & $266\ 000$ & $0.0039$ &  $84.709$ &  $12.469$ & $0.128$ &  $33.485$ &  $86.8688879$ & $8.706$ & $126.951$ \\
Titania  & $436\ 300$ & $0.0011$ & $284.400$ &  $24.614$ & $0.079$ &  $99.771$ &  $41.3514246$ & $13.46$ & $195.369$ \\
Oberon   & $583\ 500$ & $0.0014$ & $104.400$ & $283.088$ & $0.068$ & $279.771$ &  $26.7394888$ & $1.413$ & $195.37$ \\
\hline
\end{tabular}
\end{minipage}
\end{table*}
\begin{table}
\caption{Physical parameters and corresponding incertainties of the five main satellites: 
$\mathcal{G}M$ is given by \citet{jacobson07} and the mean radius of the satellites $R$ by \citet{thomas88}.}
\label{physical_parameters_satellites}
\begin{tabular}{@{}lcc@{}}
 \hline
  &$\mathcal{G}M$ & $R$\\
&$(\text{km}^3/\text{s}^2)$ &(km)\\
 \hline
Miranda& $4.4\pm 0.4 $&$235.8\pm 0.7$  \\         
Ariel&$86.4\pm 5.0$&$578.9\pm 0.6$       \\          
Umbriel&$81.5\pm 5.0$&$584.7\pm 2.8$      \\                   
Titania&$228.2\pm 5.0$&$788.9\pm1.8$\\
Oberon&$192.4\pm 7.0$&$761.4\pm2.6$\\
\hline
\end{tabular}
\end{table}

\begin{table*}
 \centering
 \begin{minipage}{140mm}
\caption{Physical parameters and corresponding incertainties of Uranus: $\mathcal{G}M$ is given by \citet{jacobson07}. The parameters $R_e$ and $R_p$ stand for 
the equatorial \citep{jacobson07} and the 
mean radius \citep{seidelman07} of the planet respectively and, $J_2$ and $J_4$ for the spherical harmonics associated with the oblateness of the planet \citep{jacobson07}.}
\label{physical_parameters_Uranus}
\begin{tabular}{@{}lccccc@{}}
\hline
&$\mathcal{G}M$& $ R_e $&$R_p$&$J_2\times 10^6$&$J_4\times 10^6$\\
&$(\text{km}^3/\text{s}^2$)&(km)&(km)&&\\
\hline
Uranus&$5\ 793\ 964\pm 6$& $26\ 200$&$25\ 362$ &$3\ 341.29\pm0.72$&$-30.44\pm1.02$\\
\hline
\end{tabular}
\end{minipage}
\end{table*}

\section{P1 : full problem}
\label{S:P1}

The full problem i.e., the N-body problem with the gravitational perturbations of each body in the system, the effect of the oblateness of the planet and the tidal effect
on the semi-major axes and eccentricities is described in this section and will be denoted by {\bf P1} in the further sections. We use a planetocentric reference frame, and consider the perturbations of $N$ satellites seen as point masses, 
and the spherical harmonics $J_2$ and $J_4$ of the gravity field of Uranus.
 In this framework, the equations of the problem are:

\begin{equation}
\vec{\ddot{r_i}}=\frac{\vec{F_i}}{m_i}-\frac{\vec{F_p}}{M}\ , 
\end{equation}
\noindent
where $\vec{r_i}=(x_i,y_i,z_i)$ locates the satellite $i$, $m_i$ being its mass and $\vec{F_i}$ the force acting on it, $\vec{F_p}$ and $M$ are 
respectively the force acting on Uranus and its mass. We write the general equations of motion for the $N$ satellites:

\begin{eqnarray}
\vec{\ddot{r_i}}&=&-\frac{\mathcal{G}(M+m_i)}{r_i^3}\nonumber\\
 &+& \sum_{j=1,j\ne i}^N \mathcal{G}m_j\left(\frac{\vec{r_j}-\vec{r_i}}{r_{ij}^3}-\frac{\vec{r_j}}{r_i^3}\right)+\mathcal{G}M\nabla_iU_i\ ,
\label{eq_Nbody}
\end{eqnarray}
with
\begin{equation}
U_i=-\sum_{n=1}^2\frac{R_e^{2n}}{r_i^{2n+1}}J_{2n}P_{2n}\left(\sin\phi_i\right)\ ,
\label{S1.Eq_potgrav}
\end{equation}
$\mathcal{G}$ being the gravitational constant, $R_e$ the equatorial radius of the planet, $\phi_i$ the latitude of the satellite $i$ in a frame connected to Uranus, and $P_n$ the classical Legendre polynomial. 
The gravitational potential (\ref{S1.Eq_potgrav}) only considers the known spherical harmonics $J_2$ and $J_4$ for the Uranian System (cfr. Table \ref{physical_parameters_Uranus}).

\par By introducing Jacobian coordinates, we can write the usual Hamiltonian to the first order on satellite masses \citep{tittemore88}:
\begin{eqnarray}
 \mathcal{H}&=&-\sum_{i=1}^{N} \frac{\mathcal{G}Mm_i}{2 a_i}\left[1+\sum_{n=1}^2 J_{2n} \bigg(\frac{R_e}{a_i}\bigg)^{2n} P_{2n} (\sin{\phi_i})\right]\nonumber \\
&-&\sum_{i<j<N}\frac{\mathcal{G}m_im_j}{a_j}\ \mathcal{R}_{ij}\ ,  
\label{S2.Hamiloniencomplet}
\end{eqnarray}
\noindent
where $a_i$ is the semi-major axis of the satellite $i$ and $\mathcal{R}_{ij}$ the disturbing function of the satellite $j$ acting on $i$. This complete Hamiltonian (\ref{S2.Hamiloniencomplet}) 
considers the mutual gravitational interactions between the $N$ satellites of the system as well as the oblateness of Uranus.

\par As dissipation effect, we add the tidal effect on the eccentricities and semi-major axes on the satellite $i$ via Kaula 
formulations (see e.g. \citealt{yoder81}):
\begin{eqnarray}
\frac{da_i}{dt}&=&3\ \frac{k_2^pn_im_iR_p^5}{Q_pa_i^4M}\ \bigg(1+\frac{51}{4}e_i^2\bigg)-21\ \frac{k_2^in_iMR_i^5}{Q_ia_i^4m_i}e_i^2\ ,\\
\frac{de_i}{dt}&=&\frac{57}{8}\ \frac{k_2^pn_im_i}{Q_pM}\bigg(\frac{R_p}{a_i}\bigg)^5\ e_i-\frac{21}{2}\ \frac{k_2^in_iM}{Q_im_i}\bigg(\frac{R_i}{a_i}\bigg)^5e_i ,
\label{eq_Kaula}
\end{eqnarray}
$R_p$ being the mean radius of Uranus $(R_p\neq R_e)$, $R_i$ and $n_i$ representing respectively the mean radius and the mean motion of the satellite $i$. 
We observe that these formulations depend on the Love number $k_2$ and on the dissipation function $Q$. These are secular equations assuming that the satellites 
are in synchronous rotation, as expected from their tidal despinning \citep{gladman96}. Moreover, due to our poor knowledge of the relevant values, the dissipation functions are assumed to be constant with respect 
to the tidal frequencies.

\section{P2 : averaged problem}
\label{S:P2}

In this section, we introduce an averaged version of the problem {\bf P1} that will be denoted by {\bf P2} in the further sections.
To reduce time computation and to elaborate an analytical perturbative theory of the problem, following \citet{tittemore88}, 
we perform an analytical averaging process on the short period angles $\lambda_i$. 
As at the lowest order, there is no coupling between the resonances in eccentricity and in inclination, 
the averaged model considers a circular-inclined approximation for the inclination resonance \citep{tittemore89} which allows us to write the Hamiltonian (\ref{S2.Hamiloniencomplet})
like:
\begin{equation}
 H=H_{kep} + H_{ob} + H_{res} + H_{sec}\ ,
\end{equation}
splitting into the keplerian, the effects of the oblateness of the planet parts, the resonant Hamiltonian, and the secular one. We obtain finally a Hamiltonian with two degrees of freedom $(J_M, J_U,\theta_M,\theta_U)$, 
these variables being canonically conjugate \citep{malhotra90}:

\begin{eqnarray}
 H&=&\nu_1 J_M+\nu_2 J_U-\beta(J_M+J_U)^2 \nonumber \\
&+&2\epsilon_4(J_M J_U)^{1/2}\cos\bigg(\frac{\theta_M-\theta_U}{2}\bigg)+2\epsilon_1 J_M\cos \theta_M \label{hamiltonianmoyenne}\\
&+&2\epsilon_2(J_M J_U)^{1/2}\cos\bigg(\frac{\theta_M+\theta_U}{2}\bigg)+2\epsilon_3 J_U\cos \theta_U \nonumber\ ,
\end{eqnarray}
\noindent
where 
\begin{eqnarray}
\theta_M & = & \theta_1\ ,\\
\theta_U & = & \theta_3\ ,\\
J_M      & = & \frac{1}{2}\ m_M [\mathcal{G}Ma_M]^{1/2}\ I_M^2 \label{eq_JM}\ ,\\
J_U      & = & \frac{1}{2}\ m_U [\mathcal{G}Ma_U]^{1/2}\ I_U^2 \label{eq_JU}\ ,
\end{eqnarray}
$\theta_i$ being the resonant angles for the resonance Miranda-Umbriel 3:1 described in Section \ref{S:key_features} and $J_i$, 
the associated variables given to the lowest order in $m_i$ and $s_i=\sin\frac{1}{2}\ I_i$. 
Following \citet{malhotra88}, we define the parameters depending on the problem:
\begin{equation}
 \nu_1=\nu_0+\Delta \nu_1 \hspace{1cm} \nu_2=\nu_0+\Delta \nu_2\ ,
\end{equation}
with 
\begin{eqnarray}
 \nu_0&=&\frac{1}{2}\bigg\{3n_U\bigg[1+3J_2\bigg(\frac{R_e}{a_U}\bigg)^2+\frac{m_M}{M}\bigg(1+\alpha\frac{d}{d\alpha}\bigg)b_{1/2}^{(0)}(\alpha)\bigg]\nonumber\\
&&-n_M\bigg[1+3J_2\bigg(\frac{R_e}{a_M}\bigg)^2-\frac{m_U}{M}\alpha^2\frac{d}{d\alpha}b_{1/2}^{(0)}(\alpha)\bigg]\bigg\}\ ,
 \end{eqnarray}
and

\begin{eqnarray}
   \Delta\nu_1&=& \bigg[\frac{3}{2}J_2\bigg(\frac{R_e}{a_M}\bigg)^2+ \frac{1}{4}\frac{m_U}{M}\alpha^2b_{3/2}^{(1)}(\alpha)\bigg] n_M\ ,\\
    \Delta\nu_2&=& \bigg[\frac{3}{2}J_2\bigg(\frac{R_e}{a_U}\bigg)^2+ \frac{1}{4}\frac{m_M}{M}\alpha b_{3/2}^{(1)}(\alpha)\bigg] n_U\ , 
\end{eqnarray}
\begin{eqnarray}
  \beta&=&\frac{3}{8}\bigg(1+9\ \frac{m_M/m_U}{\alpha}\bigg)\ \frac{1}{m_M\ a_M^2}\ . 
 \end{eqnarray}
The expression $\nu_0=0$ corresponds to the exact 3:1 commensurability between the mean motions of Miranda and Umbriel, and the $\Delta \nu_i$ are the corrections of the secular precession 
rates of the nodes $\Omega_i$ on the resonant combination of the mean motions of the satellites.
The terms $b_i^{(j)}(\alpha)$ are the Laplace coefficients, with $\alpha=a_M/a_U$. 
Their expressions have been numerically computed from the integral given by e.g. \citet{murray99}: 
\begin{equation}
b_i^j(\alpha)=\frac{1}{\pi}\int_0^{2\pi}\frac{\cos j\psi \ d\psi}{\left(1-2\alpha\cos\psi+\alpha^2\right)^i}\ .
\label{eq:laplace}
\end{equation}
We also define the expressions of $\epsilon_i$ depending on the inclination resonance $I_M^2$  \citep{malhotra90}:
\begin{eqnarray}
\epsilon_1&=&-n_M\frac{m_U}{M}\alpha f_1(\alpha)\ ,\\
\epsilon_2&=&-n_M\bigg[\frac{m_M}{M}\frac{m_U}{M}\bigg]^{1/2}\alpha^{5/4} f_2(\alpha)\ ,\\
\epsilon_3&=&-n_M\frac{m_M}{M}\alpha^{3/2} f_3(\alpha)\ ,\\
\epsilon_4&=&-n_M\bigg[\frac{m_M}{M}\frac{m_U}{M}\bigg]^{1/2}\alpha^{5/4} f_4(\alpha) \,
\end{eqnarray}
with the following expressions for $f_i(\alpha)$:
\begin{eqnarray}
    f_1(\alpha)&=&\frac{1}{8}\ \alpha \ b_{3/2}^{(2)}(\alpha)\ ,\\
    f_2(\alpha)&=&-\frac{1}{4}\ \alpha \ b_{3/2}^{(2)}(\alpha)\ ,\\
    f_3(\alpha)&=&\frac{1}{8}\ \alpha \ b_{3/2}^{(2)}(\alpha)\ ,\\
    f_4(\alpha)&=&\frac{1}{4}\ \alpha \ b_{3/2}^{(2)}(\alpha)\ ,
\end{eqnarray}
\noindent
where we neglect the indirect perturbation of order $m_M/M$.
\par To follow a trajectory through the resonance 3:1 in inclination with the tidal effects, we also consider the Kaula's formulations (\ref{eq_Kaula}) where we assume the circular approximation. 
We obtain a tidal perturbation on the semi-major axis of both satellites Miranda and Umbriel written as:
\begin{equation}
 \frac{da_i}{dt}=3\ \frac{k_2^pn_im_iR_p^5}{Q_pa_i^4M}\ , \\
 \label{eq_kaulamoyen}
\end{equation}
\noindent
where the dissipation factor of the satellites $(k_2^i/Q_i)$ is neglected in the case of the averaged problem. Because of the circular approximation, we have
a tidal perturbation only on the semi-major axes of the satellites.

\section{Numerical Methods}
\label{S:Numerical_Methods}

A major obstacle in previous studies was due to the numerical integrations of the full equations of {\bf P1} on a long enough time scale. 
The rate of the tidal evolution of the system being rather slow, the CPU time needed is quite high. 
\par Using new computing tools\footnote{The computations were performed on an HPC cluster which powers the Interuniversity Scientific Computing Facility center located at the University of Namur (Belgium). iSCF -  http://www.iscf.be.}, we have integrated numerically the equations of motion of the N-body problem in cartesian coordinates (cfr. Eq.\ref{eq_Nbody}) with 
the Adams-Bashforth-Moulton~$10^{th}$ order predictor-corrector integrator \citep{hairer08}, and the outputs have been successfully compared to the well-known software SWIFT \citep{levison94}
on the Uranian system. 
\par In this section, we introduce several methods used to obtain our numerical results. We first present the chaos indicator called Mean Exponential Growth of Nearby Orbits (MEGNO)  
\citep{cincotta00}. We perform a set of numerical integrations on the problems {\bf P1} and {\bf P2} to represent the phase plane of the problem with a chaos map. 

\par Secondly, we study the variations of the orbital elements of the satellite Miranda. According to the type of variations, different details of 
the eye of the resonance 3:1 Miranda-Umbriel, which are not visible with the chaos indicator, appear. More information is following in the subsection \ref{SS:orbital_scales}.

\subsection{Chaos indicator}
\label{SS:Chaos_indicator}
To study the stability and the structure of the phase space of our system, we implement the chaos indicator 
called Mean Exponential Growth of Nearby Orbits (MEGNO) elaborated by \citet{cincotta00} and used by different authors 
(i.e. \citealt{gozdziewski01}; \citealt{cincotta03}; \citealt{valk09}; \citealt{delsate11}). 
The indicator MEGNO is defined by: 

\begin{equation}
Y(t)=\frac{2}{t}\int_0^t\frac{\dot{\delta}(t')}{\delta(t')}\ dt'\ ,
\label{eq:megno2}
\end{equation}
where $\delta=||\vec{X}||$, $\vec{X}$ being the tangent vector $(\delta \vec{p},\delta \vec{q})$ of the vector angle action $(\vec{p},\vec{q})$ and $\dot{\delta}$, 
the time derivative of~$\delta$. Given this last equation (\ref{eq:megno2}), the running time-average of the MEGNO, $\overline{Y}(t)$, 
is defined by the following integral 
\begin{equation}
\overline{Y}(t)=\frac{1}{t}\int_0^tY(t')\ dt',
\label{eq:megno3}
\end{equation}
where, as explained by \citet{cincotta03}, $\overline{Y}(t)$ gives quickly the behaviour of the system, i.e.

\begin{eqnarray*}
 \lim_{t\rightarrow\infty}\overline{Y}(t)&=&0 \textrm{: stable and periodic trajectory}\ ,\\
 \lim_{t\rightarrow\infty}\overline{Y}(t)&=&2 \textrm{: stable and quasi-periodic trajectory}\ ,\\
\lim_{t\rightarrow\infty}\frac{\overline{Y}(t)}{t}&>&0 \textrm{:  chaotic trajectory}\ .\\
\end{eqnarray*}

\citet{gozdziewski01} propose an efficient way 
to compute the MEGNO, by adding two differential equations to the system: 

\begin{eqnarray}
\frac{dy}{dt}&=&\frac{\dot{\vec{\delta}}\cdot\vec{\delta}}{\vec{\delta}\cdot\vec{\delta}}\ t \label{gozd1}\ ,\\
\frac{dw}{dt}&=&2\ \frac{y}{t}\ ,
\label{gozd2}
\end{eqnarray}
and by computing, at each step of the integration, the MEGNO and its averaged by:

\begin{eqnarray}
Y(t)&=&2\ \frac{y(t)}{t}  \label{gozd_megno}\ ,\\
\overline{Y}(t)&=&\frac{w(t)}{t}\ .
\label{gozd_megnoaveraged}
\end{eqnarray}

In a first step, integrating the equations (\ref{gozd1}) and (\ref{gozd2}), we compute the indicator using the definitions (\ref{gozd_megno}) and (\ref{gozd_megnoaveraged}) on the problem 
{\bf P1} with Uranus and its five main satellites perturbing each other. In a second step, again on the problem {\bf P1}, we compute the indicator by considering a 3 body problem
consisting of Uranus, Miranda and Umbriel, these two satellites being involved in the mean motion resonance studied. In both cases, the first deviation vector is chosen randomly.
We notice that the main points of the global dynamics are similar in the two simulations and conclude that considering a 3 body problem is sufficient
to represent the evolution of the system during the passage in the 3:1 resonance.

\begin{figure}
\includegraphics[width=84mm]{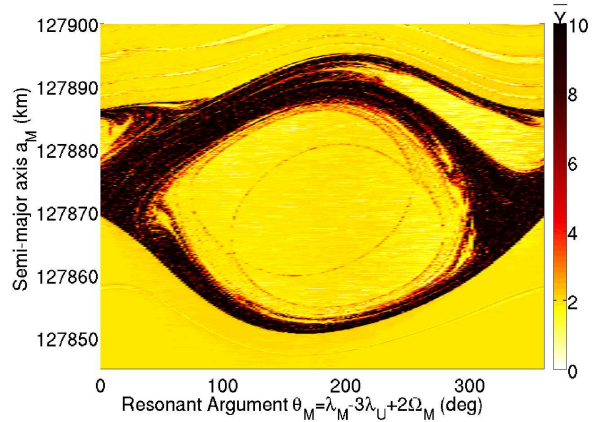}
\caption{Chaos map in the plane semi-major axis $a_M$ vs. resonant argument $\theta_M$ resulting from the numerical integrations of 
the 3 body problem Uranus-Miranda-Umbriel with the Adam-Bashforth-Moulton integrator over $1\ 500$ years. The integration step is setted to $1/80$ day. The initial conditions are fixed to the current ones 
(cfr. Tables \ref{orbital_elements_satellites}, \ref{physical_parameters_Uranus} and \ref{physical_parameters_satellites}) except 
for the mean anomaly $M_M$, the semi-major axis $a_M$ and the inclination $i_M$ of the satellite Miranda. The two first variables are set respectively between 
$[0^{\circ}-360^{\circ}[$ and $[127845\ km-127895\ km]$. The initial inclination is fixed to $4.8^{\circ}$.}
\label{fig_separatrix}
\end{figure}

To obtain the chaos map (cfr. Figure \ref{fig_separatrix}) in the plane semi-major axis $a_M$ vs. resonant argument $\theta_M$, 
we perform $10^5$ numerical integrations over $1\ 500$ years, each one associated with an initial condition of the phase plane. We fix the integration step at around $1/80^{\text{th}}$ of the smallest period of the five main satellites 
(cfr. Table \ref{orbital_elements_satellites}).
For each numerical simulation, we compute the solutions of the equations of motion of the 3 body problem {\bf P1}, without adding the tidal perturbation, and
the averaged MEGNO value (\ref{gozd_megnoaveraged}) to have the third dimension for the colour scale. The tidal effect is not considered because instead of following 
one trajectory in time, we select a region of initial conditions which covers the entire dynamics ($a_M \in [127\ 845 \ km-127\ 895\ km]$ and $M_M \in [0^{\circ}-360^{\circ}[$).
 We do not need the effect of tides to push the pair of satellites
in the resonance. This method allows us to have a global visualisation of the eye of the resonance.

\par In Figure \ref{fig_separatrix}, we obtain the phase plane given by the averaged MEGNO value. The indicator identifies the chaotic and the stable structures of the phase plane (resp. dark and light colours): 
it detects the external separatrix but also a smaller one defining the boundary of secondary resonance zones. Indeed, we 
note first the wide separatrix between the libration and the circulation zones. The stable zone in the large separatrix (top right in the figure) 
is the following primary resonance in inclination considering the nodes of Miranda $\Omega_M$ and Umbriel $\Omega_U$, represented by the resonant argument $\theta_2$. 
The other structures appearing in the circulation zones are the primary resonances 3:1 in eccentricities $\theta_i$, with $i=4,5,6$. The visualisation of 
all these structures in a small range of semi-major axis confirms the well-known narrowness between 
the resonances of the system due to the small oblateness of the planet. Secondly, we distinguish another separatrix in the libration zone 
suggesting the presence of secondary resonances zones (in the middle of the figure) that will be examined more closely in the following sections.

\subsection{Orbital elements variations }
\label{SS:orbital_scales}

The second method in our numerical analysis considers the variation in orbital elements of the satellite Miranda. 
When the bodies pass through a resonance, they cross the separatrix between the circulation and libration zones inducing changes in the nature of the orbit: we observe
modifications in eccentricities or inclinations of the satellites involved in the resonance. In contrast,
in the center of the resonance, the variations are negligible as the orbital elements are constant. Indeed, the closer is the separatrix, the stronger is the change.
We base our new colour scales, named {\it{orbital element variations}}, on these previous facts and define the following colour scales for the map:

\begin{enumerate}
 \item the variation of the semi-major axis of Miranda, $\delta_a\ ({km})$\ ,
 \item the variation of the inclination of Miranda, $\delta_i\ ({deg})$\ ,
 \item the variation of the eccentricity of Miranda, $\delta_e$\ .
\end{enumerate}

\begin{figure}
\includegraphics[width=84mm]{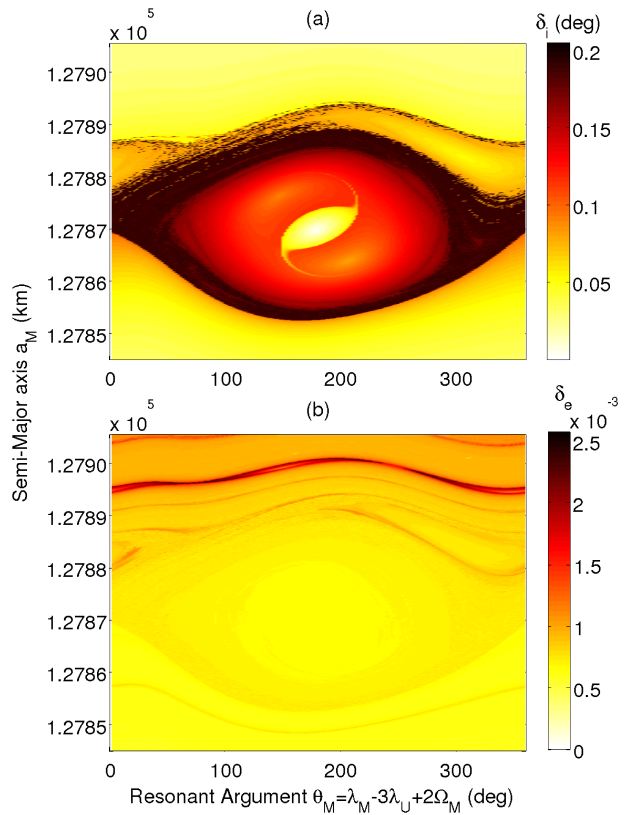}
\caption{Eyes of the resonance from the 3 body problem Uranus, Miranda, Umbriel. The integrator, the integration step, the model, 
the initial conditions are the same as Figure \ref{fig_separatrix}. The initial inclination of Miranda $i_M$ is fixed to $4.338^{\circ}$. 
The colour scales consider the variations in the inclination of Miranda $i_M (\text{deg})$ (a) or the variation in eccentricity of Miranda $e_M$ (b). 
We see the structures related to the resonance in inclination involving the node of Miranda (a) and, 
the structures related to the resonance in eccentricity involving the pericenter of Miranda $\varpi_M$ (b).}
\label{results_orbitalscale}
\end{figure}

\noindent
The study of the variations of the orbital elements allows the visualisation of structures of the space phase invisible with 
the chaos indicator MEGNO: these structures are not chaotic and/or are too barely perceptible because the variations are very small. 
We also have the association of the structures with the orbital element involved. To illustrate this, we consider the full problem {\bf P1}.
and represent the phase plane of the 3:1 resonance between Miranda and Umbriel (cfr. Figure~\ref{results_orbitalscale}) obtained by the same 
integration process than the Figure \ref{fig_separatrix}, but with the inclination (cfr. Figure \ref{results_orbitalscale} (a)) and 
the eccentricity (cfr. Figure \ref{results_orbitalscale} (b)) variations for the colour scale. What we observe is the following : we have the eye of the resonance with the inclination variation for the colour scale 
whereas it is almost erased in the eccentricity one. Moreover, with the eccentricity variation for the colour scale, we have another structure that does not appear in the 
inclination variation map: this is another primary resonance involving the pericenter of Miranda $\varpi_M$. The semi-major axis variation map brings together 
the structures of the inclination and the eccentricity variations maps giving a global visualisation of the space plane (cfr. Figure \ref{evolution_of_eye}).\\

\section{Results and comments}
\label{S:Results}

We apply the numerical methods presented in the section~\ref{S:Numerical_Methods} to the 3 body problem Uranus-Miranda-Umbriel to study the 3:1 mean motion commensurability between
Miranda and Umbriel. We consider both models {\bf P1} and {\bf P2}, the latter allowing us to speed up the CPU time while maintaining the global evolution of the system.
\par We have two different types of results. The first one groups the trajectories which evolve in the time with a tidal dissipation on a large time scale,
typically $1$\ Myr. The usual problem in this case is the dissipation function $Q$ which value is not known  but recent studies suggest smaller values for planet dissipation factor than ever (see e.g. \citealt{lainey12} 
for Saturn). We implement a set of tests with different values for $k_2/Q$ and we note similar behaviours on different time scales. 
We choose the value of $5.2\ 10^{-3}$ for the planet because it produces the rise of 
inclination of Miranda significantly higher than the current one, allowing us to study the overall process in a moderate CPU time. The strengthening of the parameters $k_2/Q$ allows us to speed up the 
integrations provided that the trajectories remain adiabatic.
\par Secondly, we represent the global evolution via maps depending on a colour scale MEGNO to represent the separatrix or depending on orbital elements variations, principally the inclination one, to study with more details the resonance. 
In this case, as a reminder, we represent the phase plane in a domain semi-major axis vs. resonant argument determined by a set of numerical integrations 
over $1\ 500$ years without taking into account the tidal effect.

\subsection{Type of results 1 : the tracking of a trajectory}
\label{SS:Tracking}

\begin{figure*}
 \vbox to220mm{\vfil
\includegraphics*[scale=0.5]{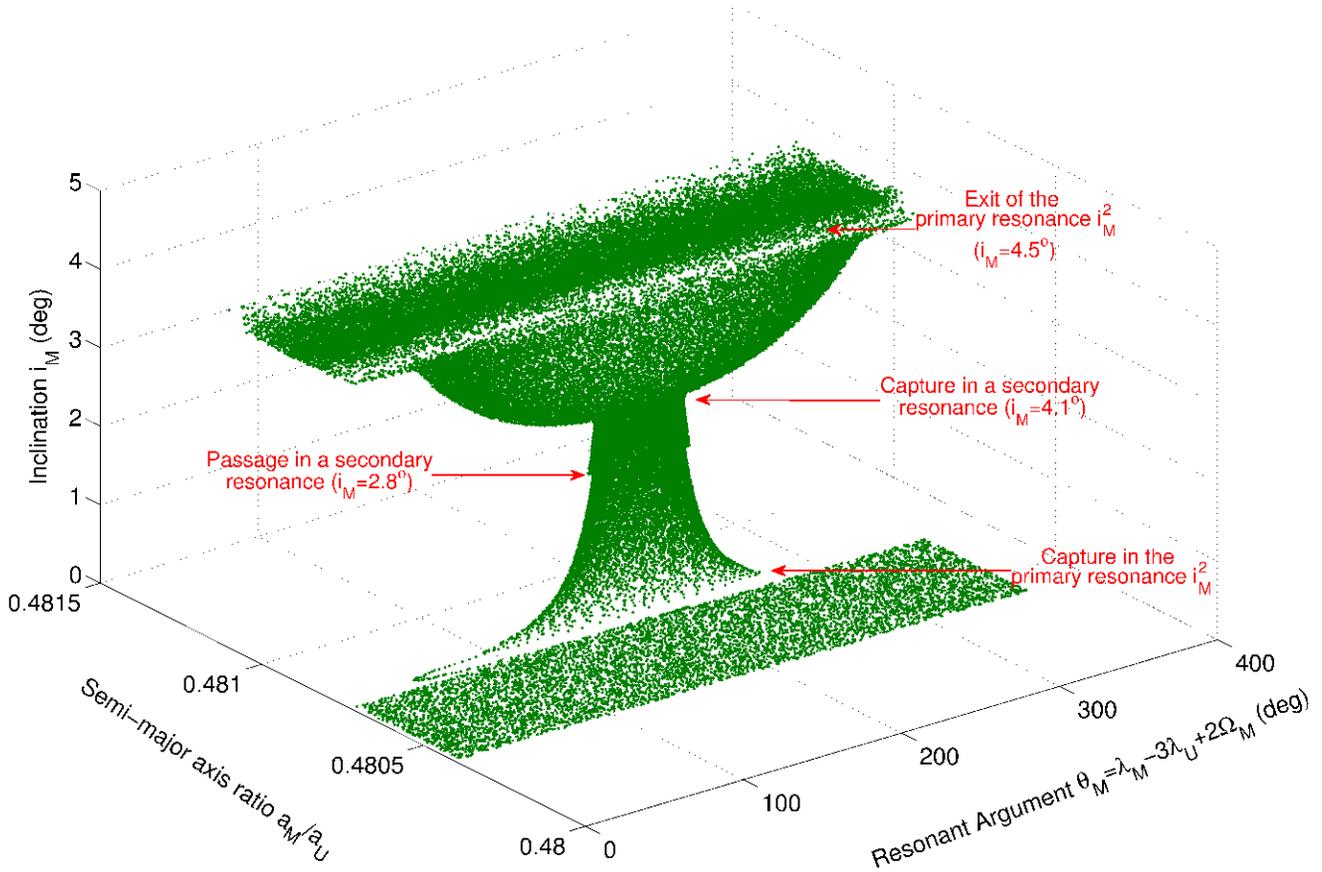}
\caption{Evolution of the resonant argument $\theta_M$ vs. the semi-major axis ratio $a_M/a_U$ vs. inclination of Miranda $i_M$. The initial conditions are the current values 
(cfr. Tables \ref{orbital_elements_satellites}, \ref{physical_parameters_Uranus} and \ref{physical_parameters_satellites}) except for 
the semi-major axis of Umbriel, $a_U=265200$ km, the semi-major axis of Miranda, $a_M=127400$ km which evolves with the tidal effect on the semi-major axes and on the eccentricities. 
The inclination of Miranda is fixed to $0.001^{\circ}$. The integration step is fixed to $1/60$ day. The ratio $k_2/Q$ is fixed to $5.2\ 10^{-3}$ for Uranus, $10^{-4}$ for the satellites.
We note the capture in the primary resonance when the commensurability $1/3$ is approached and the consequent rise of inclination of Miranda.
 We also observe the disruption of the primary resonance following the capture of a secondary resonance.}\label{figure_arbre}
\vfil} 
\end{figure*}

Let us consider the tracking  of a trajectory in its capture in the resonance 3:1 between Miranda and Umbriel. The equations of motion derive from the model {\bf P1} (cfr. Section \ref{S:P1}). 
At the begining of the simulation, the initial orbital elements are set to the current ones (cfr. Table \ref{orbital_elements_satellites}) except for the semi-major axes
of Miranda and Umbriel and the inclination of Miranda which are fixed at lower values.
We add the tidal effect on the semi-major axes and eccentricities of Miranda and Umbriel following the Kaula's formulations~(\ref{eq_Kaula}). The ratios $k_2/Q_i$ are fixed to $5.2\ 10^{-3}$ for Uranus, $10^{-4}$ for the satellites.
We can also assume a dissipation factor $Q$ higher for Miranda than Umbriel as Miranda shows an important resurfacing (\cite{smith86}) suggesting an important past thermal history
but this kind of aspect will be study soon in a following paper.
\par The result of the numerical integration gives us the evolution of the resonant argument $\theta_M $ during the capture in the resonance. In the Figure \ref{figure_arbre}, 
the tidal evolution of the resonant argument is plotted vs. the semi-major axis ratio $a_M/a_U$ vs. the rise of inclination of Miranda $i_M$. 
We observe first a circulation of the angle $\theta_M$. The capture in the resonance $I_M^2$ occurs when the commensurability $1/3$  between the semi-major axis is approaching,
corresponding to a ratio 
\begin{eqnarray*}
 \bigg(\frac{a_M}{a_U}\bigg)^{3/2}&=&(0.4807)^{3/2}\\
&=&0.3333\ .
\end{eqnarray*}
Then we enter in a libration zone and the inclination starts to increase. The trajectory is later captured in a secondary resonance zone which implies the exit of the primary 
resonance zone.
Let us note the changes of regime at the same critical points than \citet{malhotra90} ($i_M=2.1^{\circ}$, $i_M=2.8^{\circ}$ and $i_M=4.1^{\circ}$).

\subsection{Type of results 2 : Global visualisations via phase planes}
\label{SS:Global}
\subsubsection{Averaged problem {\bf P2} vs full problem {\bf P1}}

\par
\begin{figure}

\includegraphics[width=84mm]{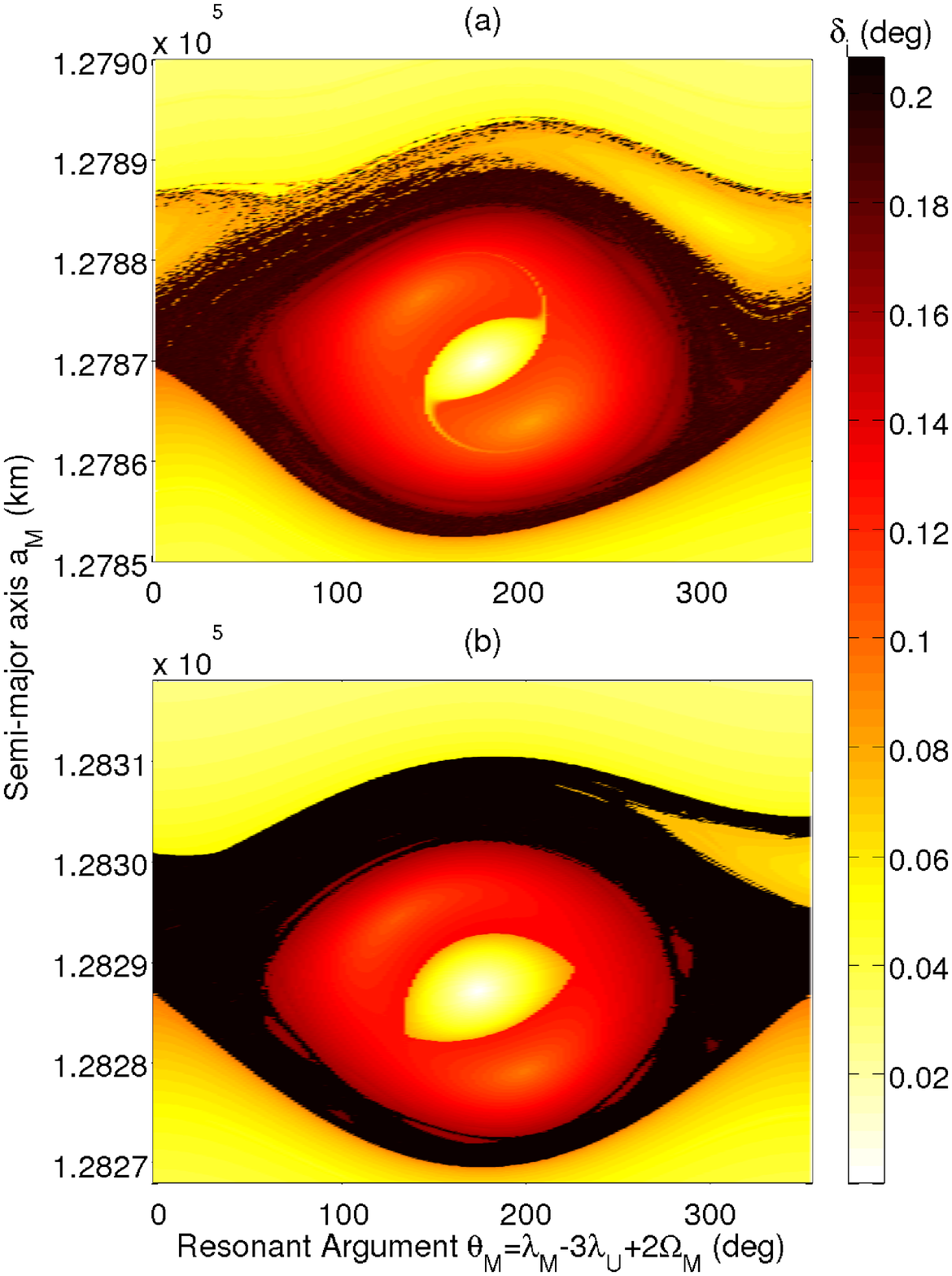}
\caption{Phase spaces from the 3 body problem Uranus, Miranda, Umbriel for the full problem {\bf P1} (a) and the averaged problem {\bf P2}.
The integration step, the model and the initial conditions are the same as Figure \ref{fig_separatrix} in the case of the Figure (a).
The initial conditions for the averaged problem {\bf P2} reflects the shift in semi-major axis $a_M$ which is set to $[128\ 268\ \text{km}-128\ 318\ \text{km}]$ and the integration step
is fixed to $17/300$ day. In both figures, the integrator is the same as Figure \ref{fig_separatrix} and the initial inclination of Miranda is $4.338^{\circ}$.
} 
\label{comparaisonMOYENCOMPLET}

\end{figure}

 Let us consider the averaged system consisting of a 3 body problem Uranus, Miranda and Umbriel in a circular-inclined
approximation described by the model {\bf P2} (Cfr. Section \ref{S:P2}).  To compare the quality of the averaged problem {\bf{P2}} vs. the full one {\bf{P1}} we implement the same process explained in the subsection \ref{SS:orbital_scales}
on the two problems. The integration step for the problem {\bf P2} is fixed to $17/300$ year, i.e. $1/300^{\text{th}}$ of the smallest nodal period. 
In our case it is the period of $17$~years related to the node of Miranda (cfr. Table \ref{orbital_elements_satellites}). The Figure \ref{comparaisonMOYENCOMPLET} compares two eyes of the resonance obtained by numerical integrations of the equations of motion of the complete model {\bf P1} for the first one (a), and
the equations of motion of the averaged model {\bf P2} for the second one (b) with the inclination variation in degrees for the colour scale.  

We observe that as a whole, the two eyes are similar: we distinguish the libration and the circulation zones, and the stable zone for the mixed resonance $\theta_2$ also appears in the two
models. We observe also the two zones of secondary resonances confirming the possibility of their studies with the averaged problem {\bf P2}. Obviously the details and the precisions of the eye in the complete model {\bf P1} are not stored in the averaged problem {\bf P2} but the global dynamics is well represented 
with the approximation.
We note that the colour scale is identical. It is due to the fact that the dynamics of the inclination is ruled by the resonant (or quasi-resonant out of the separatrix) 
argument, present in the two systems ({\bf P1} and {\bf P2}). To check our assumption we use the NAFF algorithm as Numerical Analysis of the Fundamental Frequencies based on 
Laskar's original idea (see for instance \citet{laskar93} for the method, and \citet{laskar05} for the 
convergence proofs). It aims at identifying the coefficients $a_k$ and $\omega_k$ of a complex signal $f(t)$ obtained numerically 
over a finite time span $[-T;T]$  and verifying

\begin{equation}
\label{equ:naff}
f(t) \approx \sum_{k=1}^na_k\exp(\imath\omega_kt),
\end{equation}
 where $\omega_k$ are real frequencies and $a_k$ complex coefficients. Using this tool, we check
that the variations of the inclination have the same period as this resonant angle in the two problems {\bf P1} and {\bf P2}.

\par
\begin{figure}
\includegraphics[width=84mm]{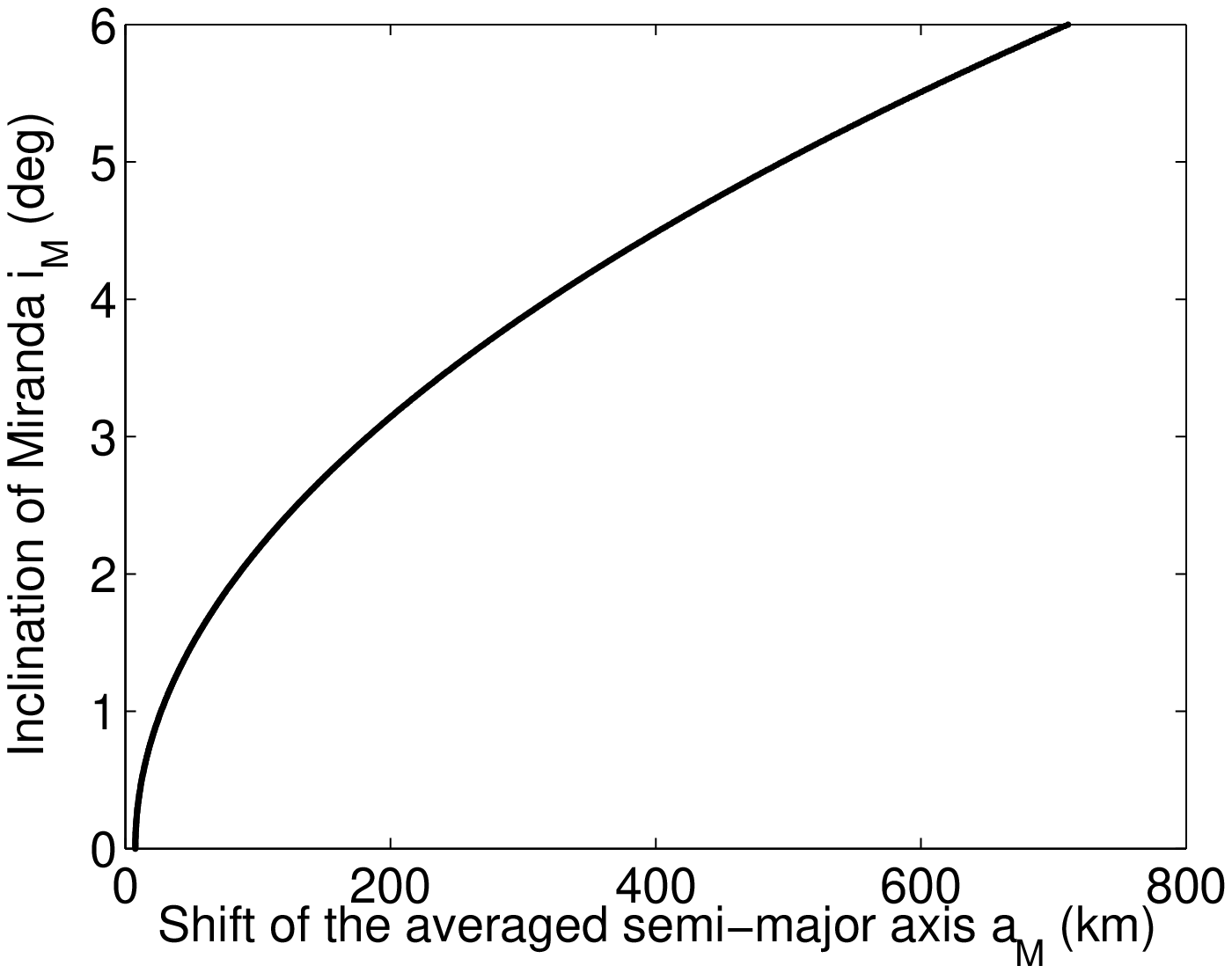}
\caption{Averaged semi-major axis of Miranda $\overline{a_M}$ vs. inclination of Miranda $i_M$ (deg). The initial condition for the semi-major axis of Miranda $a_M$ is set 
in the equation (\ref{eq_decalage}) to the center of libration in the complete model i.e. $aM=127\ 870$ km. The values of the semi-major axis of Umbriel $a_U$ and for the inclination $i_U$ are
the current ones. We see that the shift of the center of libration between the two models increases with the inclination as a power law.}
 \label{decalage}
\end{figure}

 A last thing in the comparison of the two problems is the following: looking at the two Figures \ref{comparaisonMOYENCOMPLET} (a) and (b), we notice a shift of the eye of the resonance in the two problems. Indeed the two ranges
of semi-major axis for Miranda are not the same in the two problems. This shift is due to the definitions of the variables $J_M$ (\ref{eq_JM}) and $J_U$ (\ref{eq_JU}) which imply 
that the semi-major axis of Miranda in the averaged problem differs by a quantity of order $(\sin{i_M/2})^2$ from the complete one. This can be important as the inclination increases. Based on the second chapter in \citet{malhotra88}, we show that:
\begin{equation}
 \sqrt{\overline{a_M}}=\frac{\sqrt{a_M}\ (1-3\ s_U^2)+\sqrt{a_U}\ \alpha^{-1}s_U^2}{(1-s_M^2+3s_M^2s_U^2)}\ ,
\label{eq_decalage}
\end{equation}
where $\overline{a_M}$ is the semi-major axis of Miranda in the averaged problem and $s_j=\sin{i_j/2}$, $j$ being $M$ for Miranda and, $U$ for Umbriel. The Figure \ref{decalage} 
shows the shift of the eye of the resonance with the increasing of the inclination of Miranda~$i_M$ which is significant when the inclination is high.

\subsubsection{Global evolution of the system}
The representations in three dimensions of the phase plane (semi-major axis vs. resonant argument with a colour scale) is not sufficient to represent the entire dynamics
as it evolves with the rise of inclination of Miranda. We know that the chaoticity in the 
system increases with the inclination of Miranda \citep{tittemore88}. As we want to present the whole dynamics, we propose to follow the evolution of the system
with a succesive set of maps in the semi-major axis variations for the colour scale resulting from numerical integrations of the problem {\bf P1}.

\begin{figure*}
 \vbox to220mm{\vfil
\includegraphics[scale=0.5]{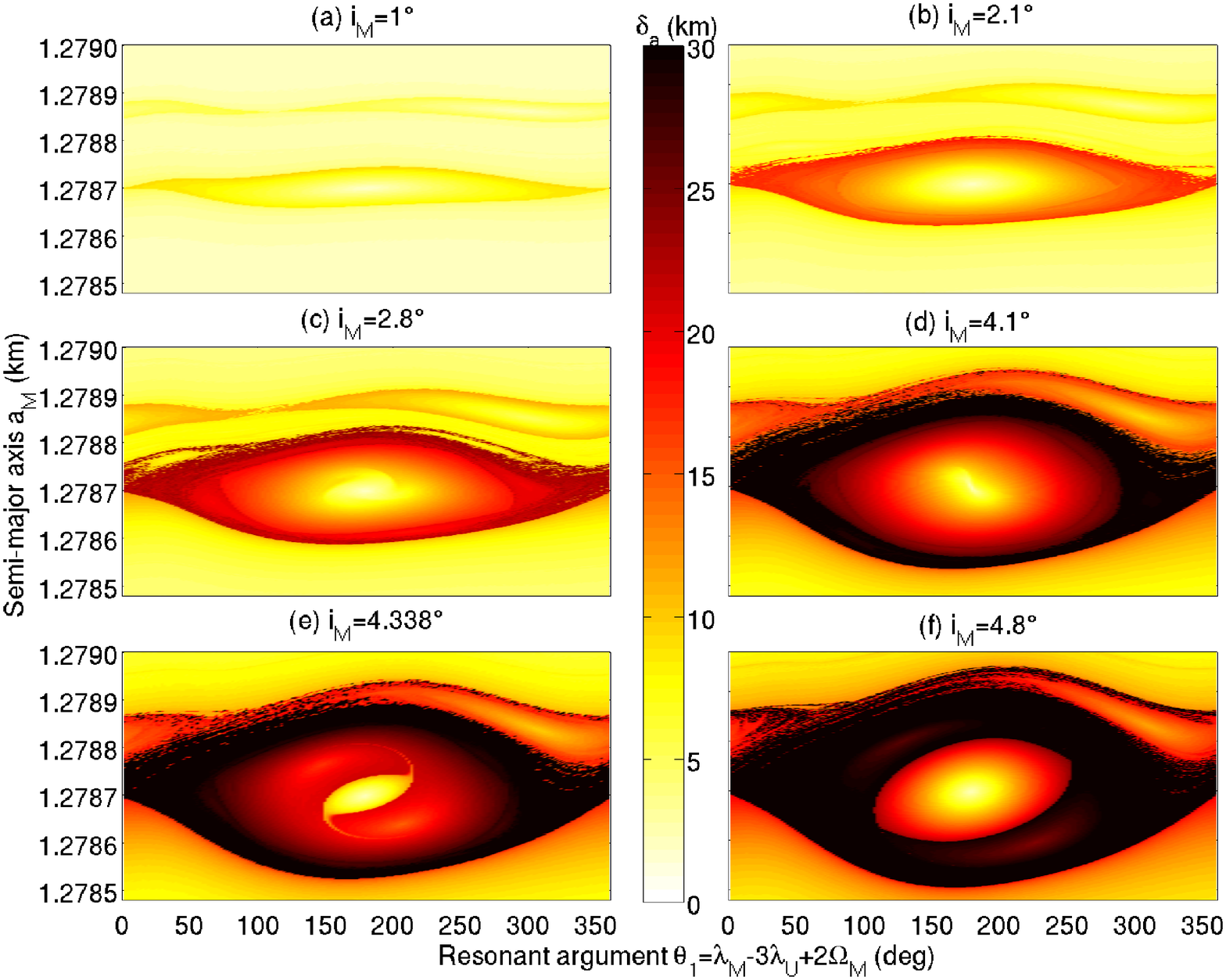}
\caption{Phase planes of the resonance from the 3 body problem Uranus, Miranda, Umbriel. The integrator, the integration step, the model, 
the initial conditions are the same as Figure \ref{fig_separatrix}. The initial inclinations of Miranda $i_M$ are fixed to $1^{\circ}$, $2.1^{\circ}$, $2.8^{\circ}$, $4.1^{\circ}$, $4.338^{\circ}$ 
and, $4.8^{\circ}$ respectively in the Figures (a), (b), (c), (d), (e) and, (f). The third dimension considers the variations in the semi-major axis of Miranda $a_M (\text{km})$. 
As the inclination of Miranda increases, the separatrix widens approaching the next primary resonances involving chaos by overlap. 
The zones of secondary resonances seem to evolve with the rise of the inclination.}
\label{evolution_of_eye}
\vfil}
\end{figure*}

The Figure \ref{evolution_of_eye} shows a set of six phase planes semi-major axis $a_M$ vs. resonant argument $\theta_1$ with the variation in semi-major
axis in colour scale. The initial inclination of Miranda is set to $1^{\circ}$, $2.1^{\circ}$, $2.8^{\circ}$, $4.1^{\circ}$, $4.338^{\circ}$ 
and, $4.8^{\circ}$ respectively. When the inclination increases, the separatrix broadens to become a layer of chaotic motion, in particular when two resonances meet.
This increase of chaos is due to the overlap of two close separatrices, consequence of the closeness of the resonances. The Figure $9$ in \citet{moons93} 
shows 'the landscape' of the problem with the location of the separatrix of the primary resonance and the centers of the secondary resonances 
in a particular plane. It also shows the location of the chaotic layers around the separatrix. We see that these layers become larger as the inclination of Miranda is high 
and we observe exactly the same feature in our maps.

\begin{figure*}
 \vbox to220mm{\vfil
\includegraphics[scale=0.5]{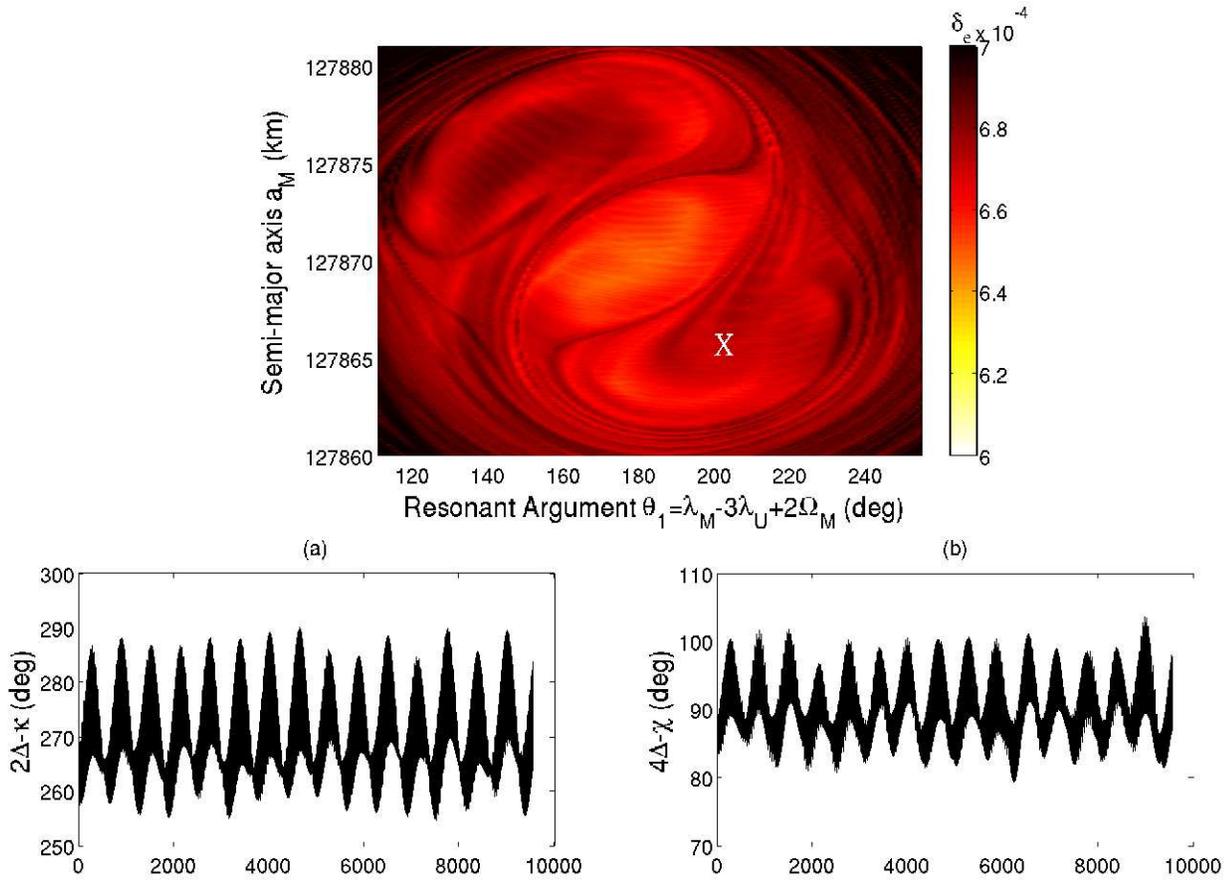}
\caption{Zoom of the center of libration from the 3 body problem Uranus, Miranda, Umbriel (Top). The integrator, the integration step, the model, 
 the initial conditions are the same as Figure \ref{fig_separatrix} except for the semi-major axis of Miranda $aM \in [127860\ \text{km}-127880\ \text{km}]$. The initial inclination of 
 Miranda $i_M$ is fixed to $4.338^{\circ}$. The symbol X in the map represents the initial condition for the trajectory analysed by frequency analysis (Below).  
 Two different combinations of frequencies seem to librate. We have the first one with the circulation frequency of $\theta_2$ (a) and 
  the second one with the circulation argument $\theta_3$ (b).}
\label{echo}
\vfil}
\end{figure*}

\par The different zones of the phase planes evolve too and, in particular, the center of libration
presents different structures moving with time: we distinguish some zones of secondary resonances which appear and move with the rise of inclination. 
Some of them have already been detected by \citet{tittemore90} and extensively studied by \citet{malhotra90} and \citet{moons93}. As we were intrigued by the different zones
in the libration center and their evolution with the rise of inclination of Miranda, we tried to make a zoom of this center. The Figure \ref{echo} shows an enlargement
of the center of the eye when the initial inclination of Miranda is set to $4.338^{\circ}$. We clearly see three zones: the center of libration and two other zones 
of secondary resonances surrounding the center.

\subsubsection{The secondary resonance zones}

\citet{dermott88} and \citet{tittemore88} show that the role of secondary resonances in the resonance 3:1 is crucial: 
a capture into a secondary resonance can explain the escape of the primary due to the chaotic layers present 
near the separatrix. To well understand the dynamics, we identify these secondary resonances via
frequency analysis and, in particular, the two zones represented in the Figure \ref{echo}. 
We select an initial condition in one of these zone (symbol 'X' in the Figure \ref{echo}) and use the resulting trajectory for the frequency analysis.
We extract the phase of the libration in the primary resonance, and combine it with the other arguments to reconstruct the
argument of the secondary resonance. We write the libration argument $\theta_1$~:
\begin{equation}
 \theta_1=\sum_i A_i \cos\bigg({\delta_i\ t +\phi_i}\bigg)\ ,
\end{equation}
 where $A_i$ and $\phi_i$ are respectively called the amplitudes and the phases of the sinusoidal function. The parameters $\delta_i$ are the libration arguments of the function.
 We extract then the principal eigenmode of the libration $\Delta$, which is associated with the bigger amplitude $A$, and combine it with the principal eigenmode of circulation of
 the arguments $\theta_2$ and $\theta_3$. Let us call them $\kappa$ and $\chi$ respectively. We obtain the result in Figure \ref{echo} where we have librations for two different combinations in a same time: 
a combination between the libration of the resonant argument $\theta_1$
\begin{itemize}
 \item  and the circulation of the argument $\theta_2$\ ,
 \item  and the circulation of the argument $\theta_3$\ , 
\end{itemize}
indicating two secondary resonances of type 2/1 and 4/1 respectively. 

\begin{figure*}
 \vbox to220mm{\vfil
\includegraphics[scale=0.5]{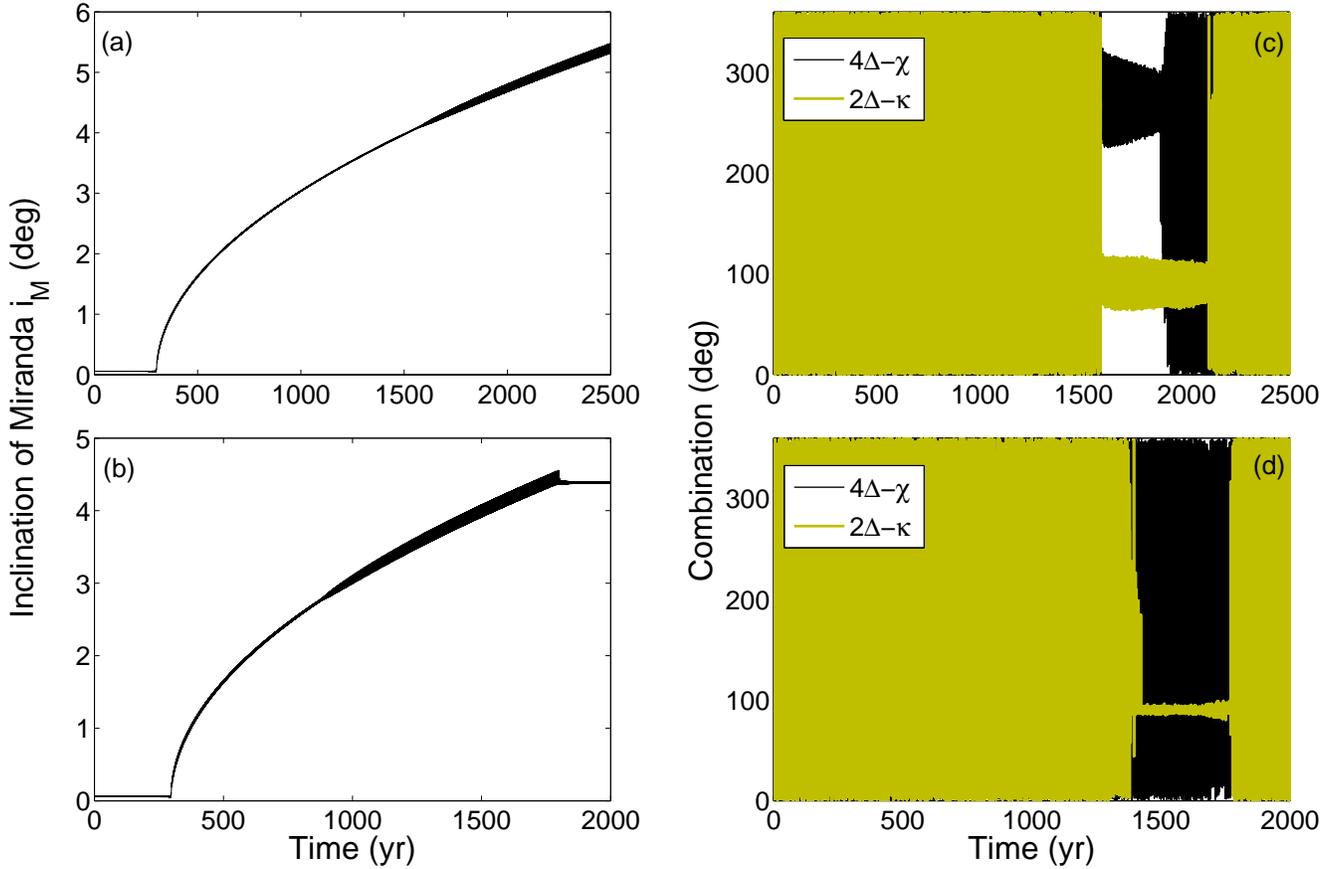}
\caption{Simulations of dissipative trajectories. The integrator, the integration step, the model, 
 the initial conditions are the same as Figure \ref{figure_arbre} except for mean longitude $\lambda_M$ of Miranda which is increased by a multiple of 9. On the left,
we have the evolution of the inclination of Miranda $i_M$ vs. time $t$ associated on the right with the two combinations $2\Delta-\kappa$ and $4\Delta-\chi$ in degrees vs. time $t$. In the first case, we observe the rise of inclination over $4.5^\circ$ (a) and the libration of the two combinations in a same time (b). 
In the second one, only the combination $2\Delta-\kappa$ is in libration (d) and we notice the exit of the primary resonance at $iM\approx 4.5^\circ$ (c).}
 \label{traj_dissi}
\vfil}
\end{figure*}

Although they are very similar, they are not the same and they present an interesting 
phenomenon as they appear in the same time. We have different elements that indicate the difference between these two secondary resonances.
First, if we change very slightly the initial conditions in the dissipative simulation, we can have one combination in a libration and the other in a circulation regime
or the two combinations in a libration regime (cfr. Fig. \ref{traj_dissi}).

More rigorously, we separate the two secondary resonances by an analytical approach using the simple model on a restricted three body problem 
described by the authors \citet{moons93}. 
The difficulty here is to have two comparable frequencies, both in a circulation regime. It is usual to use the angle-action variable to transform the
libration frequency of the primary resonance $\dot{\theta_1}$ in a circulation frequency $\dot{\Psi}$, to be compared with the circulation frequencies of the other arguments.
Following \citet{moons93}, we introduce the canonical variables:\\
\\
\begin{tabular}{ll}
 $2\sigma=-\lambda_M+3\lambda_U+2q_M$&$S=Q_M$\\
 $2\nu=\lambda_M-3\lambda_U-2q_U$&$N=2L_M-2L^\star+Q_M$\ ,
\end{tabular}
\\

\noindent where $2\sigma=-\theta_1$ and $2\nu=\theta_3$. The variables used in these definitions are the usual modified 
Delaunay's elements in case of the restricted 3 body problem ~:\\
\\
\begin{tabular}{ll}
 $\lambda_i=$\ mean longitude & $L_i=\sqrt{GMa_i}$\\
  $q_i=$\ -longitude of the node&$Q_i=L_i\ (1-\cos i_i)$\ ,\\
\end{tabular}
\\

\noindent where $i$ stands for $M$ or $U$ for Miranda or Umbriel respectively. The variable $L^\star$ is the value of $L$ at the 'exact resonance'. The Hamiltonian in these canonical variables is defined by 
 \begin{eqnarray}
  \mathcal{H}&=&\mathcal{H}_0+\mathcal{H}_1\ , \notag\\
             &=&C(N-S)^2+AN+2DS\cos2\sigma\ , \notag\\
&+&d_1i_U\sqrt{2S}\cos(\sigma-\nu)+d_2i_U\sqrt{2S}\cos(\sigma+\nu)\notag\\
&+&d_3i_U^2\cos(2\sigma)\ ,
 \end{eqnarray}
where the $d_i$ are constant coefficients defined by \citet{moons93}. Introducing the couple angle-action $(\Psi,J)$
we show that at the equilibrium $(\sigma=\pi/2$, $S=N+\frac{D}{C})$~:
\begin{equation}
 \dot{\Psi}=4D\ (1+\frac{C}{D}\ N)^{1/2}\ ,
\end{equation}
where $C,D$ are constant defined in \citet{moons93}. We compare this circulation frequency to the circulation frequency of 
\begin{eqnarray}
 \left.\frac{d}{dt} (2\nu)\right|_{\text{eq}}&=&\left.\frac{d\theta_3}{dt}\right|_{\text{eq}}\notag\ ,\\
 &=&\left.2A+4C(N-S)\ \right|_{\text{eq}}\ , \notag\\
&=&2A-4D\ ,
\end{eqnarray}
and
\begin{eqnarray}
 \left. \frac{d}{dt}\ (\nu-\sigma)\right|_{\text{eq}}&=&\left.\frac{d\theta_2 }{dt}\right|_{\text{eq}}\ ,\notag \\
&=&\left.A+4C(N-S)\ \right|_{\text{eq}}\ , \notag \\
&=&A-4D\ .
\end{eqnarray}

The expressions of the constants are given by the authors \citet{moons93} and note that the dominant term is the constant $A$. We show that the
two different frequencies are different and can thus quantitatively distinguish the two secondary resonances.

\section{Conclusions and perspectives}
\label{S:conclusionsandperspectives}
In this work, we focus on the mean motion resonance 3:1 between Miranda and Umbriel and try to explain the
high inclination of Miranda. This problem was studied by numerous authors twenty years ago but the update of 
some results with new numerical tools gives new views of it. We retrieve the main results than
these authors and improve the understanding of the problem with new powerful methods.
\par The chaos detector MEGNO has never been applied in the case
of the main satellites of Uranus. We show that the combination of the chaos detector and the orbital element maps brings a new visualisation
of the phase planes of the problem. The use of maps and frequency analysis on particular trajectories allows the detection of unusual zones
in these phase planes where two secondary resonances are combined and dominate the future evolution.
\par We show that sometimes, a trajectory could be captured in two secondary resonances in a same time: a 2/1 secondary resonance with the circulation argument of $\theta_2$ and a 4/1 secondary resonance 
with the circulation argument of $\theta_3$, the two being very close. Following the toy model of \citet{moons93}, we derive an analytical expression for a 
circulation frequency $\dot{\Psi}$ for the primary resonant argument in libration needed to compare with the other arguments in circulation.
We find two different commensurabilities and can therefore distinguish the two secondary resonances. The mix of these
two secondary resonances increases the chaos and leads to another scenario for the evolution of the system with an exit of the primary resonance
higher than $4.5^\circ$.
\par The dynamical aspects of the Uranian system are numerous and full of interest. In the future, we will study
a combination of this dynamical model with an intern evolution of the satellites. This combination is interesting by many points but the main one according to us is the
understanding of a dynamical abnormality (high inclination of Miranda) and a geological anomaly (differentiation of Miranda) through a resonance phenomenon.

\section*{Acknowledgments}
The work of EV is supported by an FNRS PhD Fellowship. The work of BN is supported by an FNRS Postdoctoral Research Fellowship.
This research used resources of the Interuniversity Scientific Computing Facility located at
the University of Namur, Belgium, which is supported by the F.R.S.-FNRS under convention
No. 2.4617.07. The authors want to thank Nicolas Delsate for his valuable advice and proofreading.

\label{lastpage}

\end{document}